%% file: WAlef_1.tex
\documentclass{evn2004}
\usepackage{txfonts}
\usepackage{natbib}
\begin{document}
   \title{A Review of VLBI Instrumentation}

   \author{W. Alef\inst{1}
          }

   \institute{Max-Planck-Institute for Radioastronomy, Auf dem H\"ugel
         69, Bonn, Germany
             }
             
             \abstract{ The history of VLBI is summarized with emphasis on the
               technical aspects. A summary of VLBI systems which are in use
               is given, and an outlook to the future of VLBI instrumentation.
             }

   \maketitle
%

   \section{Introduction}

\subsection{On the Way to VLBI}

Due to the length of radio waves the resolution of filled aperture
telescopes is much more limited in radio astronomy than in the optical
domain. Early on it was realized that in the radio higher resolution
could be achieved by using interferometers.

Michelson or phase-sensitive interferometers need relative phase-stability
between the two receiving elements. Initially radio interferometers used a
common frequency-standard, whose signal had to be transported to each
interferometer element by cable, waveguide or micro-wave link. This meant that
the separation of the receiving elements could not be expanded beyond about
100 km, as the atmospheric conditions prevent longer phase-linked baselines
via micro-wave transmission.

In the fifties of the last century in addition to conventional Michelson
interferometers so-called intensity interferometers were developed, in which
the intensity fluctuations of the incident radiation received at widely
separated antennas were cross-correlated. Due to the low requirements for
phase-stability independent local oscillators could be used and therefore
longer baselines were possible. The data could be recorded at each telescope
and correlated later. The disadvantage of this incoherent form of
interferometry is its low sensitivity and the lack of phase information which
makes it difficult to obtain images \citep{Kellermann2001}.

In the early sixties the Jodrell Bank radio linked interferometer had
reached a baseline length of up to 127 km with a resolution of better
than $1''$ at 21 cm \citep{Anderson1965}. These and other
observations showed that some radio sources were still unresolved.
Longer baselines were needed.

Discussions about the feasibility of interferometry with very long
baselines were also begun in the early sixties \citep{Matveenko1965},
the main problems being the need for independent local oscillators and
the necessity to record the data at each antenna prior to correlation.
The local oscillators had to have sufficient time stability, so that
the phase relationship between the elements of a VLB Interferometer
remained constant for some time and would allow the integration of the
correlated signals. A data recorder should allow data-rates as high
as possible while being still affordable for astronomers.

Technical developments in the middle of the sixties led to new
frequency standards like Rubidium frequency standards and Hydrogen
masers with stabilities of $10^{-11}$ and $10^{-14}$ respectively,
good enough for VLBI observations at cm wavelengths.  Relatively cheap
tape recorders for television and computers appeared on the market,
too.  This led several groups to think about developing Very Long
Baseline Interferometry systems.

\subsection{VLBI-Specific Features}

A general feature of tape recording interferometers is that the data has to be
marked with the time of its arrival at the antenna. The time marks are best
written together with the data onto the recording medium.  This will allow to
recover the arrival time of the data at the correlator. Before correlation, the
clock offsets have to be determined as precisely as possible with the help of
radio timing signals (Loran C before GPS became available) or with traveling
high-precision clocks.

A VLBI specific problem is the ``detection'' of the fringes. They can
only be found if signals from the same wavefront are correlated with
each other. The width of the fringes, or the relative time delay
between two played-back recordings which gives a detectable
interference signal, is proportional to the reciprocal bandwidth. This
means in general that the two playbacks have to be synchronized to
much better than a microsecond, in contrast to the geometrical delays
of up to several milliseconds which have to be taken into account as
well, and which are due to the figure of the earth and its motion, and
the position of the radio emitting object. All this is aggravated by
the fact that the clocks at the two antennas are independent and their
relative offsets are not known a priori.  As a result correlation has
to be done at several ``trial'' delays around the expected value to
make VLBI possible at all. 

Another difference to local interferometers is the high rate of the
fringes which can be understood as the Doppler shift between the
widely separated antennas (or the rate of change of the delay), which
is introduced by the earth's rotation. In an early paper
\citep{Matveenko1965} the authors had erroneously concluded that this
``natural'' fringe-rate would reduce the integration time to less than
a fringe period. A solution is to either remove this Doppler shift at
the stations or determine and remove the ``natural'' fringe-rate at
the correlator.

The last peculiarity is the determination of the precise delay and
delay-rate of the interference maximum after the correlation with the
so-called ``fringe fitting''.


\section{Historic VLBI systems}

In 1965 efforts to design and build a VLBI system were started by
different groups. First fringes were found in 1967 after considerable
effort. 

\subsection{The Canadian Analog VLBI System}

The system used hydrogen masers as frequency standards, and studio TV
recorders to record analog a 4 MHz band. To allow synchronization at playback
1-second ticks and spoken time marks were used. A single $2''$ tape lasted for
about 3 hours. Fringe rotation and phase-tracking correction was done at the
time of recording; adjustments were done manually.

For correlation the tapes had to be aligned manually with the
help of the audio time information. The assumption of the Canadian group that
the stability of the recorders was good enough to maintain synchronization did
not prove to be valid.  Nevertheless fringes on a 3000 km baseline were
found in spring 1967 \citep{Broten1988}.

\subsection{The MK~I Digital VLBI System}

A NRAO-Cornell group designed a digital VLBI system based on modified computer
tape drives. The 7-track tape units could record a 330 kHz band, 1-bit sampled
at a total bit-rate of 720 kbits/s. 1-bit sampling was chosen as it gives the
highest SNR in a bit-rate limited environment. For spectral line observations
where the width of the line may be small compared to the total recordable
bandwidth, 1-bit sampling is not optimal though.

A single computer tape lasted for about 150 seconds, so that 20 tapes were
needed per site and hour. The data were correlated with a general purpose
computer. To correlate one pair of tapes initially took about one hour of
computer time.  The data could easily be synchronized in the computer, and the
exact time of each bit could be recovered as long as the bit count was not
lost. First fringes were found in spring 1967 \citep{Bare1967}.

At about the same time a group at MIT-Haystack built a compatible recording
system and successfully observed OH masers with VLBI \citep{Moran1967}. Due
to the narrow line width and relative strength, OH masers have the advantage
that the precision for the time synchronization is only a few hundred
microseconds compared with the microsecond accuracy needed for continuum
sources.

Already in early 1968 the first intercontinental (global) VLBI observation
with 3 antennas (Green Bank, Haystack, Onsala) was performed. In 1970 the
maximum baseline length achievable on earth of more than 10000 km was reached.
At the same time the observing frequencies were increased from initially 610
MHz to 10.6 GHz \citep{Cohen2000,Kellermann2001}.

Geodetic VLBI observations were started in 1969. The frequency synthesis
technique was introduced to significantly increase the precision of the delay
observable -- the primary geodetic measurand. Already at that time
phase-referencing was discussed as a possibility to extend the coherent
integration time.

More multi-baseline observations were performed, and in 1971 superluminal
motion was discovered \citep{Whitney1971}. The bottleneck in VLBI observing
with the MK~I system was the correlation.  It had to be done on fairly
expensive general purpose mainframe computers.

\subsection{The MK~II VLBI system}

In about 1971 NRAO introduced the MK~II VLBI system \citep{Clark1972} which
used AMPEX $2''$ video recorders, modified to record digitized signals.
The recorded bandwidth was 2 MHz at a bit-rate of 4 Mb/s. The correlator was
implemented in hardware. The fringe rotation was done now at the correlator
because the higher observing frequencies, longer baselines, and longer
integration times required a continuous, precise, and reliable tracking of the
effects of the earth's motion. Another major argument for this change in
concept was to keep the recording site as simple as possible: correlation can
be repeated, observing not unless all stations re-observe.

By the middle of the seventies the 1-baseline correlator had been expanded to
two and three stations. In addition the number of correlation lags had been
increased from 32 to 512, which was important for spectroscopy. Initially it
was very difficult to mechanically align recording and playback machines such
that the data could be played back with reasonably low error rates. The
situation improved when $1''$ IVC recorders were introduced. 

A copy of the NRAO 3-station correlator was installed at the MPIfR in Bonn in
1978. The correlator situation further improved when Caltech built a 5 station
MK~II correlator. Later Caltech built a 16-station correlator which was then
enhanced to correlate MK~III data.

The MK~II system became a very convenient and cheap VLBI system when at the
end of the seventies VCRs replaced the IVC studio video recorders. In 1979 the
first 7-station global VLBI experiment was observed \citep{Romney1982}. At
about the same time the last $2''$ Ampex recorders were taken out of
operation.

\subsection{The MK~III VLBI system}

The MK~III VLBI system \citep{Rogers1983} was developed by MIT Haystack
Observatory in the late seventies.  Instrumentation tape recorders with 28
tracks were used to record 1-bit digitized data on $1''$ wide tape stored on
$14''$ open reels. As each single track could record up to 8 Mb/s the
maximally recordable bandwidth was a factor 56 and the sensitivity a factor of
about 7.4 larger than that of the MK~II system. Different recording modes with
lower bit-rates were possible.

Due to its 14 independently tunable baseband converters the MK~III system was
well suited for geodetic VLBI observations in which bandwidth synthesis is
used to increase the spanned bandwidth, improving
the precision of the measured delay \citep{Rogers1983,Whitney1988}. The a
priori unknown phase of the individual channels can be calibrated with the
phase calibration signals. A comb of sharp pulses with a separation of $1
\mu$s are injected in the signal path, preferentially as close to the
receiving horn as possible. In the frequency domain these pulses correspond to
sine-waves at every MHz. The injection point of the phase-cal is also the
reference point for the wide-band or multiband delay, and thus the reference
point for geodetic VLBI measurements.

In 1979 a 1-baseline correlator became operational which was soon extended so
that 3 or 6 baselines could be correlated simultaneously depending on the
recording mode. A copy of this correlator was installed at the MPIfR in Bonn
in 1982. A new and more powerful version called MK~IIIA with 6/12 baselines
capacity was developed at the end of the eighties and installed at Haystack,
USNO, and MPIfR. All MK~III correlators models had a baseline-based
architecture. In 2000 they were finally replaced by MK~IV correlators.

In 20 years of MK~III operation 3 different kinds of tape material were used.
The error rates of the original instrumentation tapes was not very good. It
was replaced when video tape on $14''$ reels became available on the market.
The tapes were 9000 feet long and could hold about 10 GB of data, and
lasted 6.5 minutes at the highest recording speed with a bit-rate of 112
Mb/s. MK~III recording was very expensive untill the introduction of movable
record/playback heads with narrow tracks ($55 \mu$m wide) increased the tape
capacity by a factor of 12 (MK~IIIA system).  The need for 24 hours of
unattended recording at the VLBA forced the introduction of the so-called thin
tapes -- a metal particle tape -- which has a length of 17600 feet and allows
a higher bit density. The capacity of the thin tapes is nearly 600 GB.

\subsection{The Canadian S2 System}

In Canada a first report on a VCR based VLBI system with a recording bit-rate
of 12 Mb/s was published in 1988 \citep{Yen1988}. Later the S2 system was
realized with professional-model VCRs recording at 16 Mb/s. By using 8 VCRs in
parallel a total bit-rate of 128 Mb/s can be recorded. A lag-based modular and
station-based correlator for the S2 system was developed \citep{Carlson1999}
for supporting both the S2-based space VLBI observations in the Japanese-led
VSOP mission and the Canadian Geodetic VLBI program. The S2 VLBI system found
widespread use also in Australia.

\subsection{Japanese VLBI Systems}

In Japan the development of VLBI systems began with the K-1 in 1976. In the
eighties the K-2, K-3, and K-4 systems followed. The K-3 system which was
developed in 1983 was fully compatible with the MK~III system, and it included
a 1-baseline correlator. The K-4 system is based on a commercial
cassette-based helical scan recorder from Sony Corporation. The initial
version used only 1-bit sampling.  It was later enhanced to multi-bit sampling
and purely digital filtering. The system can be configured for various
recording modes of which many are compatible to MK~III and VLBA modes with
bit-rates up to 256 Mb/s \citep{Kiuchi1997}. Together with a 10 station FX
correlator it supported the VSOP mission.

\subsection{Other Early VLBI developments}

\subsubsection{Real-time VLBI}

Already in 1976 a real-time link via a satellite was used to provide a
communication channel between radio telescopes in West Virginia and Ontario.
This system allowed instantaneous correlation of the data as well as a
sensitivity substantially better than that of earlier VLBI systems, because a
broader observational bandwidth could be used. \citet{YEN1977} showed that
with the use of a geostationary communications satellite it was possible to
eliminate the tape recorders and to do real-time correlation of a VLB
interferometer. As a further possibility they mention the development of a
phase-coherent VLB interferometer.

\subsubsection{Multi-view VLBI}

\citet{Hemenway1974} used four antennas at 2 sites for astrometric and
geodetic VLBI measurements. His interferometer had 2 simultaneous beams on the
sky which helps to eliminate the unknown contributions to the signal paths
caused by the ionosphere and troposphere. This observing method was later
successfully used also with 3 antenna sites and 3 to 4 telescopes per site and
the MK~III recording system \citep{Rioja2002,Porcas2003}.

The advantages of multi-beam VLBI have also been realized by the geodetic
community. By having multiple beams in different directions, parameter
correlations could be reduced and parameters of interest could be determined
more quickly and accurately \citep{Petrachenko2002}.

\subsubsection{LO Transfer via Satellite}

In 1982 a satellite was used to transfer the LO signal in a phase coherent way
between remote stations of a transcontinental VLB interferometer
\citep{Knowles1982}. Phase-degrading effects of the atmosphere were mostly
canceled out with a dual-tone transmission method. A stability of $10^{-13}$
was achieved over a 1-hour period with indications that the link was truly
phase stable at frequencies of less than 1 GHz.

\subsubsection{Near Real-time Fringe Verification}
\label{NearReal}

With the MK~III system it was possible to extract about 1 Mbit of data in the
MK~III decoder of the record terminals. This data could be transfered to one
of the correlation centers at Haystack or Bonn by modem, where it could be
correlated in software to verify the presence of fringes. As modems were very
slow in the early eighties this method was not used very much and was given up
later, as it was realized that careful checkout of the VLBI equipment before
an observations was more likely to guarantee the success of an observation.

\section{Today's VLBI systems}

\subsection{The VLBA system}

As a dedicated VLBI array with 10 equal antennas the VLBA offers reliable
performance with good amplitude calibration and easy data reduction. The VLBA
correlator can handle up to 20 stations in parallel with high spectral
resolution of up to 1024 spectral points. The VLBA system has 8 frequency
channels and can record either 1-bit or 2-bit sampled data. These differences
to the MK~III system reflect the primary drivers for the development of each
system: geodesy in the MK~III case (bandwidth synthesis) and astronomy for the
VLBA (2-bit data for spectroscopy and no need for many channels). The
correlator architecture is of an FX type; unlike most other VLBI correlators
the data is first Fourier transformed and then cross multiplied. In particular
at the time of its design the FX approach had some advantages over the XF
correlator design \citep{Romney1999}. The correlator can do pulsar gating. The
maximum total bit-rate per station is 256 Mb/s.

The VLBA recording/playback system is based on the MK~IIIA recorder, but with
completely redesigned control and record/playback electronics to improve its
reliability. 32 tracks plus 4 so-called system tracks can be written in 14
passes onto a thin tape. In order to make the system less sensitive to
individual lost tracks there is no fixed assignment between bitstream and tape
track. The assignment is changed in a round-robin fashion called
barrel-rolling. At the standard VLBA recording bit-rate of 128 Mb/s a tape
lasts nearly 12 hours, making unattended operation possible as each VLBA
station is equipped with 2 recorders. The maximum bit-rate of 256 Mb/s which
used to be only marginally higher than that of the MK~III system can now be
increased to 512 Mb/s if the 2 recorders of each VLBA telescope are scheduled
to record data in parallel.

\subsection{The MK~IV data acquisition system}

The MK~IV data acquisition system is mostly an upgrade to make MK~III more
compatible with the VLBA system. The MK~III baseband converters were equipped
with 8 and 16 MHz filters, the formatter was replaced by a MK~IV formatter
which can sample the data with 1 or 2 bits, and the recorder was modified to
read and write 32 tracks with or without barrel-rolling. Beyond those
compatibility modifications it was envisaged that the system should be able to
record 512 and even 1024 Mb/s.  The recorders have a 2nd write head for this
purpose, but they were not capable of recording reliably at 320 ips, as
required for 1024 Mb/s.  Recordings at 512 Mb/s have been done in the European
VLBI Network, but the 2-head recording mode needed for 512 Mb/s is prone
to failures.

Four MK~IV correlators became operational at the turn of the millenium. They
had been developed by a consortium of several institutions in the USA and
Europe.  The correlator is of an XF architecture. It has inputs for up to 16
stations with a maximum of 16 channels per station. The correlator is software
configurable. Cross- and auto-correlations from 32 to more than 1024 lags can
be chosen, serving both continuum and spectral-line needs.

The 3 correlators in Haystack, Washington, and Bonn operate with software
developed by Haystack Observatory. They are either exclusively dedicated to
geodetic observations like the correlator at USNO, or are used for both
astronomical and geodetic data. The correlator of the European VLBI Network at
the Joint Institute for VLBI in Europe has twice the correlator capacity
of the other MK~IV correlators. It operates with software originally developed
by Jodrell Bank personnel. It is dedicated to the correlation of
astronomical data. 

A new development at the EVN correlator allows data dumps as short as 0.25 s.
Together with many spectral channels a much wider area of the primary beam of
the single telescopes can be mapped. The aim is to map the full usable
beam-width of the Effelsberg telescope. The resulting enormous amount of raw
correlated data requires new parallel data reduction paths which are presently
being implemented.

\subsection{The Japanese Gbit VLBI System}

The first observations with the giga-bit VLBI system were performed in 1999.
The system consists of a sampler, a data recording system, and a data
correlator.  The observed data are sampled with 4-bits at a rate of 1024 Msps,
but only one bit is extracted.  The data recording system consists of a
modified commercially available high definition broadcasting recorder which
records the data at a rate of 1024 Mb/s. The correlator system was initially
developed as the real-time correlator for the Nobeyama Millimeter Array of
National Astronomical Observatory. The systems were used in a series of
geodetic and astronomical VLBI sessions. \citep{Koyama2002}

A second generation system utilizes a VLBI Standard Interface (VSI) compliant
interface. The new correlator can process two data streams at a data rate of
1024 Mbps.

\subsection{VLBI Standard Interface}

VLBI by its inherently international nature requires compatibility of recording
and playback equipment all over the world. With more and more VLBI systems
being developed which are not compatible with each other, the need for a
standard interface arose. In 1999 a draft for a
\begin{underline}V\end{underline}LBI \begin{underline}S\end{underline}tandard
\begin{underline}I\end{underline}nterface was formulated which was
finalized in 2000 (http://web.haystack.edu/vsi/). 

The purpose of VSI is to define a standard interface to and from a VLBI Data
Transmission System (DTS) that allows heterogeneous DTS's to be interfaced to
both data-acquisition and correlator systems with a minimum of effort. The
definition meanwhile has three parts: 1) a hardware definition: VSI-H 2) a
software definition: VSI-S and 3) a definition which should help transferring
VLBI via networks: VSI-E. The latter was deemed necessary if VLBI data is
transferred between different VLBI equipment at telescope and correlator via
Internet. It is still under development.

\section{Ongoing Developments}

Multi-station correlators are the most complex part of a digital VLBI system;
the data acquisition systems are kept as simple as possible to increase their
reliability.  Except for the very first system all multi-station correlators
have been realized in hardware. As a consequence the development cycle is many
years and the investment of both manpower and capital is very substantial. A
consequence is that correlators live much longer than general purpose
computers. Some MK~II correlators were operational for about 20 years. The
MK~III correlators were operated from about 1980 to 2000. The VLBA correlator
has been operational for about 10 years. It is expected that the MK~IV
correlators will not be switched off before the year 2010, even though its
design dates from the nineteen-nineties.

Rapid developments in data acquisition, transmission, and storage in the last
few years, mostly driven by the computer industry, can make VLBI cheaper and
more sensitive. Unfortunately compatibility with existing (hardware)
correlators often leads to compromises.

\subsection{Disk Recording}

In early discussions about an ``off the shelf'' replacement for the MK~IV
recorders at the end of the previous decade people considered computer
cassette tapes as a probably best choice \citep{Whitney2000}. It was soon
realized that the prices for computer hard disks (IDE/ATA) were dropping
much more rapidly than those of tape-based computer storage systems
\citep[see][]{Whitney2003}.

\subsubsection{The Mark~5 System}

In early 2001 a disk-based demonstration VLBI recording system was developed
and demonstrated within 3 months. With support from several international
institutions the Mark~5 system has been developed. The Mark~5 systems are
based on a server grade PC with a special PCI I/O card from Conduant
Corporation. This so-called Streamstor card writes to all disks in parallel in
a round-robin fashion. The data is not kept in a normal file system, and
failure of single disks does not corrupt the data on the remaining disks. The
VLBI data is fed to the Streamstor card via a FPDP bus from/to a custom design
Mark~5 I/O card, and not via the PCI bus. \citep{Whitney2003}

An initial prototype unit called Mark~5p was deployed by summer 2002. It was
limited to 512 Mb/s. Up to 16 disks were housed in individual carriers.
The Mark~5A version capable of 1024 Mb/s (with a MK~IV formatter) followed in
late 2002. Both versions were meant to be a full replacement for a MK~IV or
VLBA tape drive. For the end of 2004 the Mark~5B model has been announced
which will have a VSI compatible interface. It will make the formatter
obsolete, and will take digitized data directly from the samplers of a VLBI
terminal. It will allow the VLBA to go to 1024 Mb/s recording. In addition to
a VSI output the Mark~5B can also play back data in VLBA format. It will have
phase-cal extraction capability and for MK~IV correlators a replacement
station unit which can handle the station-based part of the correlation. Both
the Mark~5A and 5B systems can hold 2 `8-pack' disk carriers in 2 banks. Each
8-pack carries 8 disks. Recording or playback is to/from a single 8-pack at a
time. The system can record uninterrupted by switching between the two
8-packs. Idle 8-packs can be hot-swapped. Compatibility between Mark~5A and 5B
will only be realized in one direction. Mark~5A systems will be upgraded to
playback 5B recordings in VLBA format.

The Mark~5 system was chosen as the next generation VLBI recording system by
the EVN, large parts of the geodetic community, and the VLBA. The EVN and IVS
are in the process of switching to Mark~5. The VLBA will migrate to Mark~5B in
2005/2006. The EVN has officially announced the opportunity of network
observations with 1024 Mb/s, and the first astronomical gigabit observations
are scheduled for the October 2004 observing session.

\subsubsection{The PC-EVN System}

In 2001 Mujunen and Ritakari from Metsähovi Observatory presented their plan
for a disk-based VLBI recording system totally built from commodity PC
computer parts. In the original design a master PC would control 4 PCs each
recording data in parallel (RAID level 0) onto 8 removable ATA disks connected
to the PCI bus; in this configuration a recording bit-rate of 1024 Mb/s could
be reached. The data is stored on a standard Linux file system
\citep{Parsley2003}. VSI-compatible setups with single PCs are still being
used for instance in Australia for bit-rates of 256 and even 512 Mb/s.

\subsubsection{The Japanese K5 System}

A somewhat similar setup was chosen for the K5 system. A single PC will record
4 channels. The complete system with 4 PCs is compatible with the Keystone
Project Terminal and will record 16 channels \citep{Kondo2003}.

\subsection{Near Real-time Network Monitoring}

Near real-time fringe verification was successfully tried in the early
eighties (see chapter \ref{NearReal}). With the introduction of Mark~5 and
tests of e-VLBI it has been realized that the reliability of the EVN can be
greatly enhanced  if the performance of the network is checked during the
observing sessions by regularly transmitting small fractions of the data from
all participating antennas to the correlation center. In the course of 2003
`ftp fringe-tests' were introduced in the EVN. The short-time goal is to
transfer a short scan of the first observation at each frequency. Later a scan
of every observation should be transfered. At present the data is
transmitted by ftp and  correlated at JIVE with a copy of the Japanese
software correlator (see chapter \ref{JapSoft}).

\subsection{Tied Array or Real-time VLBI}

It is clear that when telescopes are connected to the Internet with a
sufficiently large bandwidth that all the data can be transferred to the
correlator directly. The cost for this transfer has to balanced against the cost of
disk storage and shipment. As long as not all telescopes have this
connectivity the data has to be recorded at the correlator so that not much is
saved. At present the development of e-VLBI activities is being pushed mostly in
Japan, where e-VLBI was first developed (see below), the European VLBI Network
and MIT Haystack Observatory.

Most research networks like e-VLBI as an application which exercises the
high-speed national and international networks which are not fully used 
at present. In particular the long-term requirements for simultaneous Gb/s
data streams from antennas worldwide converging in a single processing center
are challenging. As a result e-VLBI efforts are financially supported by
national network organizations or government funding agencies. A big stumbling
block is still the `last mile problem'. Most telescopes are in remote places
and are not connected to high-speed networks, for which optical fibers are a
pre-condition. In some countries it is required that services are bought from
a network provider which is typically very expensive, at least more expensive
than lighting a dark fiber.

From a scientist's point of view there is hardly any advantage in e-VLBI except
that the data could become available sooner after the observation.
Unfortunately some auxiliary information and measurements like that of system
temperatures and atmospheric conditions might not be available in real-time.

An exception is the UT1 campaign of the geodesists where
efforts are underway to provide UT1 within 24 hours of the observation. At
present the transfer is still of the ftp-type. As only 1 hour of
observation at 2 telescopes at a bit-rate of 112 Mb/s is involved, no special
technical problems have to be solved, and with the availability of a
sufficiently fast Internet connection at affordable annual costs the UT1
intensive observations could be changed to real-time VLBI.

Advantages of e-VLBI are mostly in savings at the logistical level and
probably in reliability after all processes will have been sufficiently
automated; manpower at antennas and correlators could be reduced, weekends and
night shifts fully utilized. 

\subsubsection{Japanese Real-time VLBI Projects}

The dominant activity in e-VLBI has been in Japan where at about 1995 the
Keystone project was started linking 4 antennas in real-time at 256
Mb/s \citep{Kondo2003}. Recently a 1 Gb/s network inside Japan is being
developed \citep{Kondo2002}.

\subsubsection{E-VLBI at MIT Haystack Observatory}

Haystack demonstrated e-VLBI between Haystack and NASA/GSFC where 788 Mb/s
were reached on a 700 km baseline in 2002. Also in 2002 real-time VLBI was
demonstrated at 256 Mb/s between Haystack and GGAO. E-VLBI test observations
were successfully conducted between Haystack and Kashima, and real-time VLBI
between Haystack and Onsala. The group is actively involved in defining the
VSI-E standard for electronic data transport for which a final definition is
expected towards the end of 2004. Research on IP protocols like RTP
for VLBI data transfer is being done.

\subsubsection{E-VLBI activities in the EVN}

The EVN has an active collaboration with several national academic network
organizations and the European network Dante/G\'eant. The immediate aim is to
demonstrate e-VLBI at 512 Mb/s with 4 or 5 telescopes at the end of
2004. The initial aim of 1 Gb/s had to be dropped because the channelization
of VLBI data in steps of powers of 2 does not match the steps of network
capacity well. So two 1 Gbit connections would be needed to transfer 1 Gb/s
VLBI data without loss. Another problem is that the speed of the present
computer main boards used in the Mark~5 systems is insufficient for dumping 1
Gb/s of data to an ethernet. Newer main board with two 1 Gbit ethernet ports
and more advanced internal architecture might solve this problem. 

In 2004 a map was made from a 3 station observation which had been observed and
correlated in real time. The bit-rate was 32 Mb/s per station and no disk
buffering was used.

\subsection{Satellite VLBI}

Ideas about satellite VLBI -- earth-space interferometers -- date back nearly
to the beginnings of VLBI. Several projects were proposed in the course of
time, but none were realized until 1997 when the Japanese VLBI satellite HALCA
was launched. A first successful experiment  was conducted however in
1986 using a 4.9-m antenna on NASA's Tracking and Data Relay Satellite System
(TDRSS).

The HALCA satellite with a 10-m antenna had 3 receivers working at 1.3, 6, and
18 cm on board. The apogee of 21,400 km and perigee of 560 km above the
Earth's surface were chosen to optimize the uv-coverage. Regular
observations were performed at 6 and 18 cm in combination with the EVN, VLBA,
and Australian radio telescopes. As the 1.3-cm receiver failed to operate
properly the maximum resolution at 6 cm of earth-space baselines is only
comparable to the VLBA alone at 2 cm, but with considerably less
sensitivity. 

\section{Future Developments}

In February of 2004 in the last general meeting of the IVS with the topic:
``Today's Results and Tomorrow's Vision'' several papers with an outlook to
the year 2010 and beyond were presented. The EVN board of directors has
started a similar discussion.

A key sentence for this chapter could be the beginning of an article by
\citet{Whitney2004}: ``In contrast to the first $\sim$30 years of VLBI
development, where highly specialized equipment for VLBI data acquisition was
designed and built at great cost, the last few years are being driven more and
more by taking advantage of rapidly developing technology in the computer and
networking industry. This trend is only likely to accelerate, and VLBI must
position itself to take maximum advantage of these technologies.''

In addition new and bigger telescopes as well as better receivers would help
to satisfy the astronomers' demand for more sensitivity. To equip telescopes
with VLBI data acquisition is becoming cheaper -- except for the obligatory
H-maser, so that additional telescopes might join the EVN. The possibility to
increase the data-rates is limited by the limited bandwidths at cm
wavelengths; mm-VLBI could profit here most.

\subsection{Data Acquisition and Transport}

In the six years between 2004 and 2010 the infrastructure in computing and
data communications will change dramatically. If the present trend continues,
in 2010 we will have 66 GHz processors with 60 GB main memory and 20 TB hard
disks.  Networks will work at 100 Gb/s speeds. Global connectivity will be
available at 660 Gb/s.  It is immediately obvious from those numbers that the
data-rate and volume problem of VLBI will have vanished. We must accept that
at least one generation (possibly two) of VLBI equipment will become totally
obsolete before we reach 2010 \citep{Mujunen2004}.  Two, four, and eight
Gb/s of recording bit-rate seems possible before 2010, both with disks and
e-VLBI.

In the VLBA and MK~IV data acquisition racks analog baseband converters (BBC)
are being used for splitting the radio band into manageable pieces, which are
then digitized and recorded. For `digital radio' frequency conversion
techniques have been developed, which encourages the design of digital BBCs,
which can offer greatly improved performance and reliability, as well as
larger bandwidths than the maximum 16 MHz of the present BBCs. The EVN has
started a Digital BBC project. The aim is to build 4 prototypes within 2 years
followed by series production. The aim is to develop a programmable,
expandable and therefore flexible converter. Initially the BBCs would be a
straight forward replacement for MK~IV and VLBA BBCs, later larger bandwidths
could be added, RFI suppression \citep[e.g.][]{Kesteven2003}, or even fringe
rotation and phase-tracking might be brought back to the antennas.  It should
be mentioned that at some point in the future it might even be possible to
design a pure software baseband converter. In Japan a first simple software
BBC with limited capabilities has already been realized.

\subsection{Correlators}
\label{JapSoft}

The correlator of the MK~I system had been realized as a program in a general
purpose computer, and in 1990 a prototype software correlator for MK~II
pulsar VLBI was devised by \citet{Petit1990}. The advances of computer
industry seem to make it possible that after more than 30 years of hardware
correlators for VLBI software correlators become again competitive. A very
attractive feature of software correlators is that the enormous effort in
manpower for a new design of the hardware does not exist; the correlator
program can simply be compiled on the latest, fast machine and voil\`a a new
correlator. A disadvantage is that at least at present hundreds and thousands
of PCs would be needed to replace a MK~IV correlator.  A software correlator
has been developed by \citet{Kondo2003} and is also being used at JIVE for the
ftp network monitoring.

For large VLBI arrays, large bandwidths, many spectral channels, and more than
1 beam, hard/firmware correlators are probably the only realistic solution for
the near and intermediate future \citep[see for example][]{Carlson2003}.
Future correlators will produce very large data-sets which will allow mapping
of a large part of the primary beam. The postprocessing software will take
over the role of stearing the delay/rate beam to the area of interest on the
sky. 

\subsection{MM-VLBI}

Higher resolution can be achieved in VLBI either by going into space to
increase the baseline length, or by going to higher observing frequencies.  At
present mm-VLBI can achieve the highest spatial resolutions, and the reduced
opacity at the higher frequencies might allow probing the cores of AGNs deeper
towards the event horizons of the central black holes.

At wavelengths of less than 1 cm the sensitivity of VLBI observations is
reduced because the telescopes are fewer and smaller, sources are often
weaker, receivers are noisier, and the maximum coherent integration time is
shorter.  As a result the number of observable sources is still limited.
Nevertheless regular mm-network observing is performed at 86 GHz with good
success (Krichbaum et al., Pagels et al., these
proceedings). 

Increasing the bit-rate is one of the keys to more sensitivity at
mm-wavelengths.  For instance a proposal by Haystack Observatory which amongst
other things asked for money for developing 4 Gb/s VLBI recording for mm-VLBI
has just been accepted by NSF.

The limited coherence time could be lengthened if attempts to measure the
path-length fluctuations in the atmosphere with water vapor radiometers are
successful (see Roy et al. these proceedings). Dual frequency receivers can be
used to calibrate the phase fluctuations at the high frequency with the lower
frequency. This technique has successfully been used in VLBI
\citep{Middelberg2002} and is also being used in the Vera project (see
Kobayashi these proceedings)

VLBI at higher frequencies is still in an experimental stage, but first
transatlantic fringes have been found even at 1.3 mm wavelength. The situation
will improve as more mm-telescopes become available at 2 and 1 mm
wavelength. (see Krichbaum et al. these proceedings) 

\subsection{E-VLA, New Mexico Array, and E-MERLIN}

When VLBI moves away from recording interferometry to become
ethernet-based the differences between VLBI and large local arrays like MERLIN
will become smaller. The difference that will (probably) remain is the use of
independent local oscillators for VLBI, while local arrays have a central LO
which is disptributed to the antennas. 

Projects like E-VLA, the New Mexico Array, and E-MERLIN with EVN extensions
demonstrate that a closer connection between VLBI on one side and the VLA and
MERLIN on the other side will develop in the future. It is even conceivable
that the VLA and the VLBA will merge into one huge array \citep{Walker2004}.
E-MERLIN may  expand to include Westerbork or Effelsberg on demand at up
to 30 Gb/s. In the far future a full integration of MERLIN and EVN might be
possible.

\subsection{Square Kilometer Array and VLBI}

The SKA will bring a major breakthrough in sensitivity in radio astronomy at
cm wavelengths. Its angular resolution though will be limited to less than
what is available today. Global VLBI with the SKA and in particular a
combination of SKA, global VLBI, and space VLBI will open a new, totally
unexplored area in the ``sensitivity -- angular resolution'' plane
\citep{Gurvits2003}. It should be noted that SKA and space VLBI alone would
not provide an appropriate coverage of the UV-plane and as a result would
compromise the gain in sensitivity by degradation in the image quality.
Telescopes paticipating in global VLBI with the SKA will feed the data to the
SKA correlator via fibres with very high data-rates.

\begin{acknowledgements}

The author would like to thank D.~A. Graham for critical suggestions and
corrections. 

\end{acknowledgements}
\bibliographystyle{aa}

\end{document}